\documentclass[aps,prl,twocolumn,preprintnumbers,superscriptaddress,showpacs,floatfix,nofootinbib]{revtex4}
\begin{document}
%\draft
%\tighten

%\twocolumn[\hsize\textwidth\columnwidth\hsize\csname@twocolumnfalse\endcsname]

\def\simlt{\stackrel{<}{{}_\sim}}
\def\simgt{\stackrel{>}{{}_\sim}}

\newcommand{\lsim}{\mbox{\raisebox{-.9ex}{~$\stackrel{\mbox{$<$}}{\sim}$~}}}
\newcommand{\gsim}{\mbox{\raisebox{-.9ex}{~$\stackrel{\mbox{$>$}}{\sim}$~}}}

\newcommand\vev[1]{{\langle {#1} \rangle}}

\renewcommand\({\left(}
\renewcommand\){\right)}
\renewcommand\[{\left[}
\renewcommand\]{\right]}

\newcommand\del{{\mbox {\boldmath $\nabla$}}}

\newcommand\eq[1]{Eq.~(\ref{#1})}
\newcommand\eqs[2]{Eqs.~(\ref{#1}) and (\ref{#2})}
\newcommand\eqss[3]{Eqs.~(\ref{#1}), (\ref{#2}), and (\ref{#3})}
\newcommand\eqsss[4]{Eqs.~(\ref{#1}), (\ref{#2}), (\ref{#3})
and (\ref{#4})}
\newcommand\eqssss[5]{Eqs.~(\ref{#1}), (\ref{#2}), (\ref{#3}),
(\ref{#4}) and (\ref{#5})}
\newcommand\eqst[2]{Eqs.~(\ref{#1})--(\ref{#2})}

\newcommand\eqref[1]{(\ref{#1})}
\newcommand\eqsref[2]{(\ref{#1}) and (\ref{#2})}
\newcommand\eqssref[3]{(\ref{#1}), (\ref{#2}), and (\ref{#3})}
\newcommand\eqsssref[4]{(\ref{#1}), (\ref{#2}), (\ref{#3})
and (\ref{#4})}
\newcommand\eqssssref[5]{(\ref{#1}), (\ref{#2}), (\ref{#3}),
(\ref{#4}) and (\ref{#5})}
\newcommand\eqstref[2]{(\ref{#1})--(\ref{#2})}

\newcommand\pa{\partial}
\newcommand\pdif[2]{\frac{\pa #1}{\pa #2}}

\newcommand\ee{\end{equation}}
\newcommand\be{\begin{equation}}
\newcommand\eea{\end{eqnarray}}
\newcommand\bea{\begin{eqnarray}}

\newcommand\mpl{M_{\rm P}}

\newcommand\dbibitem[1]{\bibitem{#1}\hspace{1cm}#1\hspace{1cm}}
\newcommand{\dlabel}[1]{\label{#1} \ \ \ \ \ \ \ \ #1\ \ \ \ \ \ \ \ }
\def\dcite#1{[#1]}

\def\calf{{\cal F}}
\def\calh{{\cal H}}
\def\call{{\cal L}}
\def\calm{{\cal M}}
\def\caln{{\cal N}}
\def\calp{{\mathcal P}}
\def\calr{{\cal R}}
\def\calpr{{\calp_\calr}}

\newcommand\bfa{{\mathbf a}}
\newcommand\bfb{{\mathbf b}}
\newcommand\bfc{{\mathbf c}}
\newcommand\bfd{{\mathbf d}}
\newcommand\bfe{{\mathbf e}}
\newcommand\bff{{\mathbf f}}
\newcommand\bfg{{\mathbf g}}
\newcommand\bfh{{\mathbf h}}
\newcommand\bfi{{\mathbf i}}
\newcommand\bfj{{\mathbf j}}
\newcommand\bfk{{\mathbf k}}
\newcommand\bfl{{\mathbf l}}
\newcommand\bfm{{\mathbf m}}
\newcommand\bfn{{\mathbf n}}
\newcommand\bfo{{\mathbf o}}
\newcommand\bfp{{\mathbf p}}
\newcommand\bfq{{\mathbf q}}
\newcommand\bfr{{\mathbf r}}
\newcommand\bfs{{\mathbf s}}
\newcommand\bft{{\mathbf t}}
\newcommand\bfu{{\mathbf u}}
\newcommand\bfv{{\mathbf v}}
\newcommand\bfw{{\mathbf w}}
\newcommand\bfx{{\mathbf x}}
\newcommand\bfy{{\mathbf y}}
\newcommand\bfz{{\mathbf z}}

%units
\newcommand\yr{\,\mbox{yr}}
\newcommand\sunit{\,\mbox{s}}
\newcommand\munit{\,\mbox{m}}
\newcommand\wunit{\,\mbox{W}}
\newcommand\Kunit{\,\mbox{K}}
\newcommand\muK{\,\mu\mbox{K}}

\newcommand\metres{\,\mbox{meters}}
\newcommand\mm{\,\mbox{mm}}
\newcommand\cm{\,\mbox{cm}}
\newcommand\km{\,\mbox{km}}
\newcommand\kg{\,\mbox{kg}}
\newcommand\TeV{\,\mbox{TeV}}
\newcommand\GeV{\,\mbox{GeV}}
\newcommand\MeV{\,\mbox{MeV}}
\newcommand\keV{\,\mbox{keV}}
\newcommand\eV{\,\mbox{eV}}
\newcommand\Mpc{\,\mbox{Mpc}}

\newcommand\msun{M_\odot}

\newcommand\sub[1]{_{\rm #1}}
\newcommand\su[1]{^{\rm #1}}

\newcommand{\one}{_1}
\newcommand{\two}{_2}

\newcommand\mone{^{-1}}
\newcommand\mtwo{^{-2}}
\newcommand\mthree{^{-3}}
\newcommand\mfour{^{-4}}
\newcommand\mhalf{^{-1/2}}
\newcommand\half{^{1/2}}
\newcommand\threehalf{^{3/2}}
\newcommand\mthreehalf{^{-3/2}}

\renewcommand{\P}{{\cal P}}
\newcommand{\f}{f_{\rm NL}}
\newcommand{\mk}{{\mathbf k}}
\newcommand{\mq}{{\mathbf q}}

\newcommand{\zetag}{{\zeta\sub g}}
\newcommand\sigmas{{\sigma^2}}

\newcommand{\fnl}{{f\sub{NL}}}
\newcommand{\fnltilde}{{\tilde f\sub{NL}}}
\newcommand{\fnli}{ {f_{ {\rm NL}i } } }

\newcommand\kmax{{k\sub{max}}}
\newcommand\bfkp{{{\bfk}'}}
\newcommand\bfkpp{{{\bfk}''}}
\newcommand\bfpp{{{\bfp}'}}

\newcommand\tp{{(2\pi)}}
\newcommand\tpq{{(2\pi)^3}}
\newcommand\tps{{(2\pi)^6}}

\renewcommand\ni{{N_{,i}}}
\newcommand\nj{{N_{,j}}}
\newcommand\nij{{N_{,ij}}}
\newcommand\nii{{N_{,ii}}}

\title{The inflationary prediction for primordial non-gaussianity}

\author{David H. Lyth}
\email{d.lyth@lancaster.ac.uk}
\affiliation{Department of Physics, Lancaster University, Lancaster LA1 4YB, UK}
\author{Yeinzon Rodr\'{\i}guez}
\email{y.rodriguezgarcia@lancaster.ac.uk}
\affiliation{Department of Physics, Lancaster University, Lancaster LA1 4YB, UK}
\affiliation{Centro de Investigaciones, Universidad Antonio Nari\~no, Cll 58A \# 37-94, Bogot\'a D.C., Colombia}
 
%\date{\today}

\begin{abstract}
We extend the $\delta N$ formalism so that it gives all of the stochastic  
properties of the primordial curvature perturbation $\zeta$ if the initial field perturbations
are gaussian. The calculation requires only the knowledge of some family of unperturbed universes.
A formula is given for the normalisation $\fnl$ of the bispectrum of $\zeta$,
which is the main signal of non-gaussianity. Examples of the use of the 
formula are given, and its relation to cosmological perturbation theory
is explained.
%clarified.togehits
%We give examples, including the first-time calculation of the parameter $f_{NL}$,
%which gives the level of non-gaussianity, in a modular (two-component) inflation model,
%the reproduction of the $f_{NL}$ parameter in the curvaton scenario, and the correction of
%another.
%$\zeta$ calculated in a hybrid-type (two-component) model of inflation; the latter two examples being originally calculated using
%a  known result from second-order perturbation theory.
\end{abstract}

\maketitle

{\em Introduction.}~
The primordial curvature perturbation  of the Universe, 
denoted here by $\zeta$, is already present a few
Hubble times before cosmological scales start to enter the horizon
\cite{treview}.
 Its time-independent value at that
stage seems to set the initial condition for the subsequent evolution of
all cosmological perturbations. As a result, observation probes the stochastic
properties of $\zeta$, which
is found to be almost gaussian  with an almost
scale-invariant spectrum.

According to present ideas $\zeta$ is  supposed
to originate from the vacuum fluctuations
 during inflation of one or more light 
scalar fields, which on each scale are promoted to classical perturbations
around the time of horizon exit. 
One takes 
inflation to be  almost exponential (quasi de-Sitter spacetime)
corresponding to a practically constant
Hubble parameter $H_*$, and  the effective masses of the fields
to be  much less than $H_*$. This ensures that the
fields are almost massless and live in almost unperturbed quasi de-Sitter spacetime, making their 
%field perturbations are
perturbations
indeed almost gaussian and scale-invariant. 
This automatically makes $\zeta$ almost scale-invariant, and can
(though not automatically \cite{luw,lr})  make it also almost gaussian.

All of this is  of intense interest at the present time,
because  observation over  the next few years  will rule out 
most existing scenarios for the generation of $\zeta$, by detecting or
bounding the scale-dependence and  non-gaussianity of $\zeta$. In this 
Letter we describe a general  procedure for calculating
 the level of non-gaussianity, by means of the
$\delta N$ formalism \cite{ss,sasaki1}.

{\it Defining the curvature perturbation.}~Perturbations 
of the observable Universe are defined with respect to an unperturbed reference
universe, which is homogeneous and isotropic.
Its line element may be written as
\mbox{$ds^2 = -dt^2 + a^2(t) \delta_{ij} dx^i dx^j$}
defining the unperturbed scale factor $a(t)$, time $t$, and the Cartesian
spatial coordinates $\bfx$.

The curvature perturbation is only of interest after the universe has
been smoothed on some scale $\left(\frac{k}{a}\right)^{-1}$ much bigger than the horizon
$H^{-1}$.
To define it, one
%considers  spatial slices of the Universe
%with uniform energy density. Their 
%metric can be written as \cite{lr,sb,zc,sasaki1}
takes the fixed-$t$ slices of spacetime to have uniform energy density, and the fixed-$x$
worldlines to be comoving. The spatial metric is \cite{lr,sasaki1,sb,zc}
\be
g_{ij} = a^2(t) e^{2\zeta(t,\bfx)} \gamma_{ij}(t,\bfx) =\tilde a^2(t,\bfx)\gamma_{ij}(t,\bfx)
\,.
\label{zetadef}
\ee
In this expression, $\gamma_{ij}(t,\bfx)$  has unit determinant, so that
a  volume of the Universe bounded by fixed spatial coordinates
 is proportional to the locally-defined scale
factor  $\tilde a (t,\bfx)$. In the inflationary scenario the factor
$\gamma_{ij}$  just accounts for the tensor perturbation, but its form is
irrelevant here. 
According to this definition, $\zeta$ is the perturbation
in $\ln\tilde a$.
%Only the spatial variation of $\zeta$ is significant, and to make contact
%with observation we can work with its Fourier components in a box
%a  bit bigger than the observable Universe, setting the zero mode
%equal to zero so that $\zeta$ has vanishing spatial average.

One can also
consider a slicing 
whose metric has the form in Eq. \eqref{zetadef} without the $\zeta$
factor, which we call  the flat slicing.
Starting from any initial flat slice at time $t\sub{in}$, 
let us define the amount of expansion
$N(t,\bfx)
\equiv \ln\left[\frac{\tilde a(t)}{a(t\sub{in})}\right]$
to a final slice of uniform energy density.
 Then \cite{ss,sasaki1}
\be
\zeta(t,\bfx) = \delta N \equiv N(t,\bfx) - N_0(t)
\,,
\label{deln1}
\ee
where  $N_0(t)\equiv \ln\left[\frac{a(t)}{a(t\sub{in})}\right]$
is the unperturbed amount of expansion.

To make use of the above  formalism we assume
that in the super-horizon regime ($aH\gg k$), the evolution of the 
Universe at each position (the local evolution)
is  well-approximated 
by the evolution of  some unperturbed universe \cite{sasaki1,llmw,lyth}.
This `separate universe' assumption
%must  be correct on sufficiently large scales,
%because in that limit it can be derived from the exact Einstein equations
%by dropping spatial gradients \cite{sasaki1}. 
%On a given scale, it will be correct 
%throughout 
%the super-horizon regime provided that the given scale is the biggest relevant
%one. In particular, it
will presumably be correct on cosmological scales
because these scales are so big \cite{lyth}.

By virtue of the separate universe assumption, $N(t,{\bf x})$ is the amount of expansion in some
unperturbed universe, allowing $\zeta$ to be evaluated knowing the evolution of a family of such universes.
For a given content of the Universe it can be checked using the gradient expansion \cite{sb,sasaki1,rs,lv} method, but we do not
wish to assume a specific content.

%An immediate consequence of the separate universe assumption is that the 
% worldlines comoving with  the flow of energy
%  are  practically orthogonal to the slices of uniform energy
%density (because that is the case in an unperturbed universe).
%Therefore the uniform density slicing practically coincides with the 
%one orthogonal to the comoving worldlines, and  can equally well be used
%to define $\zeta$ as it is implicitly assumed by many authors.
%Another consequence is that 
The separate universe assumption leads also to
local energy conservation,
so that $\zeta$ is conserved as long as the pressure is a unique function of the energy density.
%Indeed,  using the uniform-density
%slicing, and remembering that $\tilde a$
%determines the expansion,
%\be
%\dot\rho(t)= -3\tilde H(\rho(t) + P(t,{\bf x}))= -3\left(H+\dot\zeta\right)(\rho(t)+P(t,{\bf x}))
%\,,
%\label{rhocons}
%\ee
%where $\tilde H\equiv \frac{\dot{\tilde a}}{\tilde a}$,
%$\rho$ is the energy density, and $P$ is the pressure. 
%During any era when 
%$P$ is a unique function of $\rho$ (the adiabatic condition),
%$P$ is uniform on the chosen slicing; then $\dot\zeta$ 
%vanishes (because it is uniform and its
%spatial average vanishes) so that $\zeta$ is conserved.
This consequence of the separate universe assumption was
 first recognised in
full generality in Refs. \cite{sasaki1,rs} 
(see also Ref.~\cite{sb} for the case of inflation,
Refs. \cite{lyth,llmw} for the case of linear perturbation theory, and
Ref. \cite{lv} for a coordinate-free treatment).

{\em The inflationary prediction.}~The 
evolution of the observable Universe, smoothed on the shortest cosmological
scale, is supposed to be 
determined by the values of one or more light scalar fields when that scale
first emerges from the quantum regime a few 
Hubble times after horizon exit. %\footnote
%{In principle heavy fields could also be relevant, but they are generally
%supposed to be fixed during inflation at the instantaneous minimum of the 
%potential determined by the light fields; in other words they are supposed
%to be integrated out of the action. There is also the tensor perturbation
%which does not affect our results.}.
Defined on a flat slicing, each
field $\phi_i$ at this epoch
will be of the form $\phi_i(\bfx) = \phi_i + \delta\phi_i(\bfx)$.

Because quasi exponential inflation is assumed, and only light fields are considered,
%it is a good approximation to take the $\delta\phi_i$ to be massless fields
%living in unperturbed quasi de-Sitter spacetime \cite{treview}. In these circumstances
the  perturbations $\delta\phi_i$ generated from the vacuum are 
 almost gaussian,  with an almost
 flat spectrum
 %\hbox{$\calp_{\delta \phi_i} = \left(\frac{H_*}{2\pi}\right)^2$ \cite{bd}}. 
 %where $H_*$ is the
 %almost-constant Hubble parameter
 \cite{bd}
\be
\calp_{\delta \phi_i} = \left(\frac{H_*}{2\pi}\right)^2 \,. \label{phispec}
\ee 

Now we invoke the separate universe assumption, 
and  choose the homogeneous quantities $\phi_i$ to correspond to the 
unperturbed universe. Then \eq{deln1} for  
$\zeta$ becomes 
\bea
\zeta(t,\bfx) &=& N(\rho(t),\phi_1(\bfx),\phi_2(\bfx),\cdots)
- N(\rho(t),\phi_1,\phi_2,\cdots)
\label{master0}
\,. \nonumber \\
&&
\eea
In this  expression, the expansion $N$ is evaluated in an unperturbed 
universe, from an epoch when the fields have assigned values to one
when the energy density has an assigned value $\rho$.
This expression \cite{ss,sasaki1} allows one to 
propagate forward the  stochastic properties of $\zeta$
to the epoch when it becomes observable, given those of  the initial field 
perturbations.

Since the observed curvature perturbation is almost gaussian,
it must be given to good accuracy by one or more of the linear terms
(we use the notation $\ni\equiv \frac{\partial N}{\partial\phi_i}$ and
\mbox{$\nij\equiv \frac{\partial^2 N}{\partial\phi_i\partial\phi_j}$}) 
\be
\zeta(t,\bfx) \simeq  \sum_i  \ni(t)   \delta\phi_i (\bfx)
\,,
\label{linearterm}
\ee 
with the field
perturbations being almost gaussian.
%\footnote{
%Here,  $\ni\equiv \frac{\partial N}{\partial\phi_i}$ and
%$\nij\equiv \frac{\partial^2 N}{\partial\phi_i\partial\phi_j}$.}.
In this Letter we include for the first time the quadratic terms
%\footnote{
%(Here and elsewhere, we  are not
%displaying a  homogeneous term needed to make the spatial average
%of $\zeta$ vanish.)
\bea
\zeta(t,\bfx) \simeq \sum_i  \ni(t)  \delta\phi_i 
+ \frac12\sum_{ij} \nij(t) \delta\phi_i \delta\phi_j
\,.
\label{master1}
\eea
They may be
 entirely  responsible for any observed non-gaussianity
if  the field perturbations are gaussian to sufficient
accuracy.
%(Here and elsewhere, we  are not
%displaying a  homogeneous term needed to make the spatial average
%of $\zeta$ vanish.)

{\em The bispectrum.}~The stochastic properties of the perturbations
are specified  through expectation values 
which, according to the inflationary paradigm, are taken with respect to
the time-independent (Heisenberg picture) quantum state of the 
Universe
%\footnote{
(to be precise, the quantum state of the universe
 before it somehow collapses to give the 
observed Universe).
Focusing on $\zeta$, we consider
 Fourier components, 
 \hbox{$\zeta_\bfk\equiv
 \int d^3k \zeta(t,\bfx)\exp(i\bfk\cdot\bfx)$}.

The stochastic properties of a gaussian perturbation are specified entirely by the spectrum $\calp_\zeta$,
defined through 
$\vev{\zeta_\bfk\zeta_\bfkp} \equiv (2\pi)^3 P_\zeta(k) \delta^3(\bfk+\bfk')$
and $\calp_\zeta(k) \equiv \frac{k^3}{2\pi^2} P_\zeta(k)$.
From \eqs{phispec}{linearterm}
\be
\calp_\zeta  = \left(\frac{H_*}{2\pi}\right)^2 \sum_i \ni^2
\label{zspec}
\,.
\ee

Non-gaussianity is defined through higher correlations. We consider only the three-point correlation. (The four-point correlation may give a competitive observational signature and can be calculated in a similar fashion \cite{bl,trispec}.) It defines the  bispectrum $B_\zeta$ through
$\vev{\zeta_\bfk\zeta_\bfkp\zeta_\bfkpp} \equiv \tp^3 B_\zeta(k,k',k'') 
\delta^3(\bfk+\bfkp+\bfkpp)$. Its  normalisation is
 specified by
a parameter $\fnl$ according to \cite{spergel,maldacena}
\be
B_\zeta \equiv -\frac{6}{5}\fnl(k,k',k'') ( P_\zeta(k) P_\zeta(k') + 
{\rm cyclic})
\,.
\label{fnldef}
\ee
(In first-order cosmological perturbation the gauge-invariant gravitational potential $\Phi$
during matter domination before horizon entry is $\Phi = -\frac53\zeta$, and our definition of
$\fnl$ coincides with the definition \cite{spergel}
\mbox{$B_{\Phi} \equiv 2 \fnl(k,k',k'') [ P_\Phi(k) P_\Phi(k') + 
{\rm cyclic}]$}.
At second-order these definitions of $\fnl$ differ \cite{lastbartolo}.)

We shall take $\calp_\zeta$ and $\fnl$ to be evaluated when 
cosmological scales approach the horizon and $\zeta$
becomes observable. Observation gives  $\calp_\zeta=(5\times 10^{-5})^2$,
and $|\fnl|\lsim 100$ \cite{komatsu}. Absent a 
 detection, this will eventually come down to roughly $|\fnl|\lsim 1$ \cite{spergel}.

Ignoring any non-gaussianity of the $\delta\phi_i$, our
 formula in Eq. \eqref{master1} 
makes $\fnl$ practically 
independent of the wavenumbers.
Indeed, generalising the result found in Ref. \cite{bl}, we have calculated
\be
-\frac35 \fnl =
\frac{\sum_{ij}\ni\nj\nij}{2\[ \sum_i \ni^2 \]^2 }
+ \ln(kL) \frac{\calp_\zeta}  2 
\frac{\sum_{ijk} \nij N_{,jk} N_{,ki}}{\[ \sum_i \ni^2  \]^3}
\,. \label{master2}
\ee
In deriving  this expression we used the spectrum $\left(\frac{H_*}{2\pi}\right)^2$
of the field perturbations, and used \eq{zspec} to eliminate
 $H_*$ in favour of 
 $\calp_\zeta$.
As discussed in Ref. \cite{bl}, the  logarithm 
can be taken to be of order 1, because
it involves the size $k\mone$  of a typical scale under consideration,
relative to the  size $L$ of the region  
within which  the stochastic properties are specified.
Except for the logarithm, {\em $\fnl$ is scale-independent if the 
field perturbations are gaussian}.

If only one $\delta\phi_i$ is relevant,
 \eq{master1}  becomes
\be
\zeta(t,\bfx) = \ni \delta\phi_i + \frac12 \nii (\delta\phi_i)^2
\label{simple1}
\,,
\ee
and because the first term dominates, \eq{master2} becomes
\be
-\frac35\fnl = \frac12 \frac{\nii}{\ni^2}
\label{simple2}
\,.
\ee
In this case, $\fnl$ may equivalently be defined  \cite{spergel} 
by writing $\zeta = \zeta\sub g - \frac 35 \fnl \zeta\sub g^2$,
where $\zeta\sub g$ is  gaussian.

To include 
 the possible non-gaussianity of the $\delta\phi_i$, 
we define the bispectra $B_{ijk}$ of the dimensionless field perturbations
$(2\pi/H_*)\delta\phi_i$  and their normalisation $f_{ijk}$, 
in exactly  the same way that we defined $B_\zeta$ and $\fnl$.
%Remembering that the field perturbations  are almost gaussian
%corresponding to $|f_{ijk}|\ll 1$, their non-gaussianity adds to
%$\fnl$  a contribution 
These bispectra add the following contribution to $\fnl$ in Eq. (\ref{master2})
\be
\Delta \fnl = \frac{\sum_{ijk} \ni \nj N_{,k} f_{ijk}(k,k',k'') }
{\( \sum_i \ni^2 \)^{3/2} } \calp_\zeta^{-1/2}
\label{delfnl}
\,.
\ee
The   $f_{ijk}$,
 generated directly from the vacuum fluctuation,  will 
depend strongly on the wavenumbers.

{\em Cosmological perturbation theory.}~In the super-horizon regime
the non-linear theory \cite{sasaki1}
that we have used is a complete description.
The basic expression in \eq{master0} is non-perturbative, giving $\zeta(t,\bfx)$ in
terms of the initial fields and the expansion of a family of unperturbed
universes. The  second-order expansion in \eq{master1} is  a matter of convenience.
As  we shall see it seems to be adequate in practice, but \eq{master0} would still
be applicable if the expansion converged slowly or not at all.
%(In this connection, it is worth remarking that the calculation of the 
%trispectrum, which might be another signal of non-gaussianity \cite{trispec},
%could require the inclusion of more terms even in the situations that we shall
%encounter.)

Cosmological 
perturbation theory (CPT) is completely different.
It is applicable both inside and outside the horizon, being at each instant 
 a power series in the 
perturbations  of the metric and the stress-energy tensor, together with
whatever variables are needed to completely specify the latter and close
the system of equations. During inflation these variables are the components
of the inflaton, while afterwards they may involve oscillating fields
and a description of the particle content. First-order CPT
 is usually adequate and
%as we shall see in the curvaton example,
can describe non-gaussianity at the level $\fnl \gg 1$, which has to be
generated by the second-order term in Eq. \ref{master1}.
%as it is described in Ref. \cite{luw} for the curvaton scenario,
%is {\em not} necessarily equivalent to the truncation of \eq{master1} at first order
%in order to calculate $\fnl$.
Second order CPT is generally needed only 
to handle non-gaussianity at the level $|\fnl|\sim 1$.

Quantised CPT is needed to calculate the stochastic properties of the
%spectrum and bispectrum of the
initial field perturbations $\delta\phi_i$, 
which are the input for our calculation.
The slow-roll spectrum in \eq{phispec} comes from the first-order
calculation. The bispectrum is a second-order effect 
and has, in the context of slow-roll inflation, been calculated
in Refs.~\cite{maldacena,seery2}. It is shown elsewhere \cite{ignacio} that
$|\Delta \fnl| \ll 1$ in this case.
%In this case $\zeta$ is time independent and our formalism is of no
%interest. It is found that $|\fnli|$ is of order $\zeta$
%times slow-roll parameters, making it negligible. Hopefully a similar result
%can be established more generally (a first attempt, but still in the context
%of single component inflation, is made in \cite{seery}),
%and from now on we take the $\delta\phi_i$
%to be gaussian (one example which differs appreciably from this expectation
%is Ghost inflation \cite{ghost} where the peculiar form of the Lagrangian leads to
%a highly non-gaussian inflaton field).
Higher correlators have not been calculated yet
%The trispectrum, which is a third-order effect, has not been calculated yet
and would give an additional contribution to Eq. (\ref{master2}) which presumably
%is
is also negligible.
Exotic non slow-roll models \cite{gdk} can make $|\Delta \fnl| \gg 1$, but from now on
we set $\Delta \fnl = 0$.

In the regime $aH \gg k$, perturbation theory must be compatible with Eq. (\ref{master1}).
In particular, the non-local terms, present at second order for a generic perturbation,
must be absent for $\zeta$ (see also Ref. \cite{rst}). Finally,
CPT is needed to evolve the perturbations after horizon
entry, but that is not our concern here. 
%This Letter is concerned with the era between horizon exit and horizon
%entry, where  CPT is in principle made redundant by our non-linear formalism.
In the following, we apply our formalism to calculate
$\fnl$ in various cases and  compare it with  the  CPT result where that 
is known.

{\em A two-component inflation model.}~As a first use of \eq{master2} we 
consider the two-component inflation model of Kadota and Stewart \cite{ks},
estimating for the first time  the non-gaussianity which it predicts.
The model
 works with a complex field $\Phi$, which is supposed to be a modulus
with a point of enhanced symmetry at the origin. Writing
$\Phi \equiv |\Phi|e^{i\theta}$, the tree-level potential has a maximum at
$\Phi=0$ and depends on both $|\Phi|$ and $\theta$. A one-loop correction
turns the maximum into a crater and inflation occurs while $\Phi$
is rolling away from the rim of the crater. The curvature 
perturbation is supposed to be constant after the end of slow-roll 
inflation. For  $\theta\ll\theta_c$, with $\theta_c$ being a parameter of the model,
it is found that $N\propto \left|\frac{\theta_c}{\theta}\right|$. Through the first term of \eq{master2}
$\fnl\simeq \left|\frac{\theta}{\theta_c}\right|$ which is too small ever to be observed.

{\em The curvaton model.}~ In the curvaton model \cite{curvaton} (see also Ref. \cite{earlier})
the curvature perturbation $\zeta$ grows, from a negligible value in an initially radiation dominated epoch, due to the oscillations of a light field $\sigma$ (the curvaton) around the minimum of its quadratic potential $V_\sigma(t,\bfx) = \frac{1}{2}m_\sigma^2 \sigma^2(t,\bfx)$, where $m_\sigma$ is the curvaton effective mass. Due to the oscillations, the initially negligible curvaton energy density redshifts as \mbox{$\rho_\sigma (t,\bfx) \approx \frac{1}{2}m_\sigma^2 \sigma_a^2(t,\bfx) \propto a^{-3}(t,\bfx)$}, where $\sigma_a$ represents the amplitude of the oscillations. Meanwhile the radiation energy density $\rho_r$ redshifts as $a^{-4}$. Soon after the curvaton decay, the standard Hot Big-Bang is recovered and $\zeta$ is assumed to be conserved until horizon reentry.

To calculate $\fnl$ using Eq. (\ref{master2}) we first realise that $\sigma_*$ (the value of $\sigma$ a few Hubble times after horizon exit) is the only relevant quantity since the curvature perturbation produced by the inflaton, and imprinted in the radiation fluid during the reheating process, is supposed to be negligible. Thus, \eq{simple2} applies. Second, we can redefine $N$ as the number of e-folds from the beginning of the sinusoidal oscillations to the curvaton decay. This is because the number of e-folds from the end of inflation to the beginning of the oscillations is completely unperturbed as the radiation energy density dominates during that time. Thus, $N$ is now a function of three variables
\be
N(\rho_{dec},\rho_{osc},\sigma_*)=\frac{1}{3}\ln\left(\frac{\rho_{\sigma_{osc}}}{\rho_{\sigma_{dec}}}\right) = \frac{1}{3}\ln\left[\frac{\frac{1}{2}m_\sigma^2 [g(\sigma_*)]^2}{\rho_{\sigma_{dec}}}\right]
\,,
\ee
where $g \equiv \sigma_{osc}$ is the amplitude at the beginning of the sinusoidal oscillations, as a function of
%the value of $\sigma$ soon after horizon exit.
%its value $\sigma_*$ a few Hubble times after horizon exit.
$\sigma_*$. Here the curvaton energy density just before the curvaton decay $\rho_{\sigma_{dec}}$ is expressed in terms of the total energy density $\rho_{dec}$ at that time, the total energy density at the beginning of the sinusoidal oscillations $\rho_{osc}$, and $g$ by
%\be
$\rho_{\sigma_{dec}} = \frac{1}{2}m_\sigma^2 [g(\sigma_*)]^2 \left(\frac{\rho_{dec} - \rho_{\sigma_{dec}}}{\rho_{osc}}\right)^{3/4}$.
%\,.
%\ee
%we define instead $N$ as a function of the radiation and curvaton energy densities $\rho_{r_{osc}}$ and $\rho_{\sigma_{osc}}$ at the beginning of the curvaton oscillations, and the uniform total energy density $\rho_{dec}$ just before the curvaton decay. The curvature perturbation, now given by $\zeta = \frac{\partial N}{\partial \rho_{\sigma_{osc}}} \delta \rho_{\sigma_{osc}} + \frac{1}{2}\frac{\partial^2 N}{\partial \rho_{\sigma_{osc}}^2} (\delta \rho_{\sigma_{osc}})^2$,
After evaluating $\frac{\partial}{\partial \sigma_*} = g' \frac{\partial}{\partial g}$, at fixed $\rho_{dec}$ and $\rho_{osc}$, we obtain 
\be
N_{,\sigma_\ast} = \frac{2}{3} r \frac{g'}{g} \,,
\ee
where $r \equiv \frac{3\rho_{\sigma_{dec}}}{3\rho_{\sigma_{dec}} + 4\rho_{r_{dec}}}$
being $\rho_{r_{dec}}$
the radiation energy density just before the curvaton decay, so that
\be
\mathcal{P}_\zeta = \frac{H_\ast}{2\pi} N_{,\sigma_\ast} = \frac{H_\ast r}{3\pi} \frac{g'}{g} \,,
\ee
in agreement with first-order cosmological perturbation theory \cite{luw}. Differentiating again we find
from \eq{simple2}
\be
\fnl = -\frac{5}{6} \frac{N,_{\sigma_* \sigma_*}}{N,_{\sigma_*}^2} = \frac{5}{3} + \frac{5}{6} r - \frac{5}{4r} \left(1 + \frac{gg''}{g'^2}\right)
\,,
\ee
%is then expressed in terms of the quantities
%where $r = \frac{3\rho_{\sigma_{dec}}}{3\rho_{\sigma_{dec}} + 4\rho_{r_{dec}}}$, being $\rho_{r_{dec}}$ the radiation energy density just before the curvaton decay,
%and $g$ (the amplitude of $\sigma$ at the beginning of the oscillations as a function of its value a few Hubble times after horizon exit: $\sigma_{osc} = g(\sigma_*)$)
%by
%\be
%\zeta = \zeta^{(1)} + \frac{1}{2}\left[- 2 - r + \frac{3}{2r}\left(1 + \frac{gg''}{g'^2}\right)\right]\left(\zeta^{(1)}\right)^2 \,,
%\ee
%after evaluating the derivatives of $N$ with respect to $\rho_{\sigma_{osc}}$. The $\fnl$ parameter, as defined in Eq. (\ref{master2}), is thus given by\footnote{Primes mean derivatives with respect to $\sigma_*$.}
%\be
%\fnl = \frac{5}{3} + \frac{5}{6} r - \frac{5}{4r} \left(1 + \frac{gg''}{g'^2}\right) \,,
%\ee
which nicely agrees with the already calculated $\fnl$ using first- and second-order perturbation theory (see Refs. \cite{luw,lr,bartoloc}).

{\em Another two-component  model.}~Finally  we consider
the  two-component inflation model of Refs. \cite{lr,ev}. 
For at least some number $N$ of
$e$-folds after cosmological scales leave the horizon, the 
 potential is
%\be
\mbox{$V = V_0 \( 1 + \frac12 \eta_\phi \frac{\phi^2}{m_P^2} + \frac12 \eta_\sigma \frac{\sigma^2}{m_P^2} \)$},
%\,,
%\label{evv}
%\ee
with the first term dominating, $\eta_\phi$ and $\eta_\sigma$ being
the slow-roll $\eta$ parameters, and $m_P$ being the reduced Planck mass. The idea is to use \eq{master2} to calculate
the non-gaussianity after the  $N$ $e$-folds which, barring 
cancellations, will  place  a lower limit on the observed non-gaussianity.

The  slow-roll equations give the field values $\phi(N)$ and $\sigma(N)$, 
in terms of
those obtaining just after horizon exit;
$\phi(N) = \phi \exp(-N\eta_\phi)$ and $\sigma(N) = \sigma
 \exp(-N\eta_\sigma)$.
This gives  $V(N,\phi,\sigma)$ and allows us to calculate
the derivatives of $N$ with respect to $\phi$ and $\sigma$ at fixed $V$.
Focusing on the  case $\sigma=0$ considered in Ref. \cite{ev}, we find
\be
\zeta = \frac{\delta\phi}{\eta_\phi \phi} - \frac{\eta_\phi}{2}
\( \frac{\delta\phi}{\eta_\phi \phi} \)^2 +  \frac{\eta_\sigma}{2}
e^{2N(\eta_\phi - \eta_\sigma)} \( \frac{\delta\sigma}{\eta_\phi \phi} \)^2
\,, \label{finalzetaev}
\ee
%(This method of calculating $\zeta$ clearly generalises to any separable potential.)
%This result agrees with the second-order perturbation theory equation presented in Ref. \cite{malik1}.
%It disagrees with the second order equation presented in Ref. \cite{ev} (based on Ref. \cite{acquaviva})
%through a single non-local term \cite{evlr}.
in agreement with the second-order perturbation calculation of Ref. \cite{malik1}. 
If the observed  $\zeta$ has a non-gaussian part $\zeta_\sigma$
equal to the last term of Eq. (\ref{finalzetaev}) and a gaussian part generated mostly
{\em after} inflation, one can obtain $|\fnl| >1$ by choosing
$\eta_\phi > 0.26$, $\eta_\sigma = \frac{\eta_\phi}{2}$, $N=70$, and $\zeta_\sigma = 10^{-2} \zeta$.

This model was studied originally \cite{lr,ev} using a second-order 
perturbation expression for the {\em time-derivative} of $\varepsilon H m_P^2 \zeta_\sigma$, with $\varepsilon$
being the $\varepsilon$ slow-roll parameter.
This expression disagrees with ours \cite{newlr} through the appearance of non-local terms, though the order of magnitude is similar \cite{antti}.

{\em Acknowledgments.}~In connection with this work, 
D.H.L. has benefited particularly from correspondence with Misao Sasaki.
D.H.L. is  supported by PPARC grants Nos. PPA/G/O/2002/00469,  PPA/V/S/2003/00104,
PPA/G/S/2002/00098, and PPA/Y/S/2002/00272, and by European Union Grant No.
MRTN-CT-2004-503369.  Y.R. is fully supported by COLCIENCIAS (COLOMBIA), 
and partially supported by COLFUTURO (COLOMBIA), UNIVERSITIES UK (UK), 
and the Department of Physics of Lancaster University (UK).

\end{document}